\documentclass[conference]{IEEEtran}
\usepackage{amssymb,amsmath,amsfonts}% need for subequations, \align
\usepackage{cite}                    % use for citing multiple references within the same brackets
\usepackage{graphicx}
\usepackage{hyperref}
\usepackage{suffix}
\usepackage{mathtools}

\ifCLASSINFOpdf
\else
\fi
\begin{document}
\title{Secrecy Analysis and Learning-based Optimization of Cooperative NOMA SWIPT Systems}

\author{\IEEEauthorblockN{Furqan Jameel\IEEEauthorrefmark{1}, Wali Ullah Khan\IEEEauthorrefmark{2}, Zheng Chang\IEEEauthorrefmark{1}, Tapani Ristaniemi\IEEEauthorrefmark{1}, Ju Liu\IEEEauthorrefmark{2}} 
\IEEEauthorblockA{\IEEEauthorblockA{\IEEEauthorrefmark{1}Faculty of Information Technology, University of Jyv\"askyl\"a, FI-40014 Jyv\"askyl\"a, Finland.\\} \IEEEauthorrefmark{2}School of Information Science and Engineering, Shandong University, Qingdao, People's Republic of China.\\ 
	}}
	
\maketitle

\begin{abstract}
Non-orthogonal multiple access (NOMA) is considered to be one of the best candidates for future networks due to its ability to serve multiple users using the same resource block. Although early studies have focused on transmission reliability and energy efficiency, recent works are considering cooperation among the nodes. The cooperative NOMA techniques allow the user with a better channel (near user) to act as a relay between the source and the user experiencing poor channel (far user). This paper considers the link security aspect of energy harvesting cooperative NOMA users. In particular, the near user applies the decode-and-forward (DF) protocol for relaying the message of the source node to the far user in the presence of an eavesdropper. Moreover, we consider that all the devices use power-splitting architecture for energy harvesting and information decoding. We derive the analytical expression of intercept probability. Next, we employ deep learning based optimization to find the optimal power allocation factor. The results show the robustness and superiority of deep learning optimization over conventional iterative search algorithm.  
\end{abstract}


\section{Introduction}

Non-orthogonal multiple access (NOMA) has received much hype due to its promise to effectively utilize the wireless spectrum. NOMA works by allowing users to share the same temporal/ spatial resources while the receiving side carries out successive interference cancellation (SIC) \cite{8648865,8644111}. On the other hand, cooperative communications can help by improving the system capacity, extend the coverage area and achieve a higher degree of freedom with single antenna nodes. Thus, the idea of user cooperation in NOMA has attracted much interest due to its applications in 5G and has given birth to an important research topic called the cooperative NOMA. It was first proposed in \cite{7117391} wherein, a user with the stronger channel decodes the message and then assist by relaying the message to the far NOMA user.   

Despite substantial improvements in terms of spectral efficiency, the research work on energy efficient cooperative NOMA schemes is still at infancy stage. To that end, simultaneous wireless information and power transfer (SWIPT) has drawn much research interest due to the ability of RF signals to transfer information and energy at the receiver \cite{diamantoulakis2017joint}. Thus, applications of SWIPT in NOMA have been studied from the perspective of outage performance, cooperation, and energy harvesting (EH) efficiency \cite{ding2016relay}. However, owing to the dual function of RF signal and broadcast nature of NOMA, the transmission from source to destination can be eavesdropped by a malicious user. More specifically, the EH receivers can intercept the confidential information being exchanged between legitimate users. In order to provide security to the low-powered devices, physical layer security (PLS) has been introduced as an alternative to computation heavy cryptographic techniques \cite{8103768}. PLS techniques can improve the secrecy performance of wireless networks by means of cooperative relaying, jamming and multiple-antenna beamforming.

In \cite{liu2016cooperative}, the authors proposed energy and spectral efficient protocol by combining NOMA with SWIPT. They showed that the proposed scheme does not jeopardize the diversity gain of the edge users while enabling the cell-center users to self-power themselves. In \cite{diamantoulakis2017joint}, Diamantoulakis \emph{et al.} consider downlink and uplink multiple access protocols for SWIPT systems. In particular, they investigate the performance in the downlink for NOMA and time division multiple access (TDMA), while for uplink conditions they consider NOMA with time sharing. These works were extended for multiple-input-single-output (MISO) NOMA for hybrid time switching and power-splitting SWIPT architecture in \cite{do2018improving}. They also derived tight closed-form expressions of the outage probability and demonstrated the superiority of cooperative NOMA over conventional NOMA and OMA systems. Besides these developments, few investigations have been conducted for improving the secrecy performance of NOMA using SWIPT. The authors of \cite{8171177} maximized the secrecy sum rate by optimizing the allocated power. Closed-form expression of optimal power-splitting ratio was derived and it was shown that the proposed method outperforms the uniform power allocation methodology. In \cite{8333737}, Zhou \emph{et al}. proposed a cooperative jamming technique for energy harvesting multiple-input-single-output (MISO)-NOMA cognitive radio systems. The authors claimed that the proposed scheme for NOMA outperforms the conventional orthogonal multiple access (OMA) scheme in terms of power efficiency. 

Of late, deep learning has emerged as a key technique for improving the performance of wireless networks. Deep learning is a part of machine learning consisting of multiple hidden layers \cite{lecun2015deep}. More specifically, in contrast to shallow machine learning methods, deep learning has multiple intermediate layers of neurons between input and output layers. At each hidden layer, the weighted sum of the previous layers are updated and an activation function is applied \cite{nair2010rectified}. The authors of \cite{hinton2006fast} first proposed the idea that deep learning is an important and powerful tool for handling non-linear and complex problems. Some other works considered deep learning for the physical layer, multiple-input-multiple-output (MIMO) systems, and channel coding \cite{gruber2017deep,wang2017deep}. This positive trend also attracted much-needed attention to multiple access schemes. Thus, the authors of \cite{kim2018deep} optimized the sparse code multiple access (SCMA) scheme using deep learning. To do so, they developed a strategy for selecting the codebook which minimizes the bit error rate (BER) while using the minimum amount of computation time. Another important study that integrates orthogonal frequency division multiplexing (OFDM) and deep learning was conducted by the authors of \cite{ye2018power}. It was shown that the deep learning approach performs best for signal detection and channel estimation. More recently, the authors of \cite{gui2018deep} used long short-term memory (a branch of supervised deep learning) for data detection in uplink NOMA. They showed that the deep learning based NOMA scheme is more reliable as compared to conventional hard-decision optimization solutions. 

So far, it has become evident that the work on secrecy performance of energy harvesting cooperative NOMA systems is very limited. Moreover, the work on deep learning approaches for physical layer security of NOMA is non-existent. Therefore, in order to advance this promising field of wireless communications, we consider a scenario where energy harvesting cooperative NOMA users communicate in the presence of an energy harvesting eavesdropper. We derive the analytical expression of intercept probability of DF energy harvesting cooperative NOMA system which, according to the authors' best knowledge, has not been derived in the literature. We also optimize the secrecy performance of the system by using deep learning for finding the optimal value of the power allocation factor. The results of deep learning approaches are then compared with benchmark iterative search algorithms. It has been shown that the deep learning based NOMA scheme is robust and computationally lightweight.    

The remainder of the paper is organized as follows. Section II provides details of the system model. In Section III, the analytical results for intercept probability are provided. In Section IV, deep learning based neural network model is discussed. Section V provides numerical results and their relevant discussion. In Section VI, some concluding remarks are provided. 

%%%%%%%%%%%%%%%%%%%%%%%%%%%%%%%%%%%%%%%%%%

\section{System Model}
\begin{figure}
    \centering
        \includegraphics[trim={0 6cm 0 0},clip,width=0.45\textwidth]{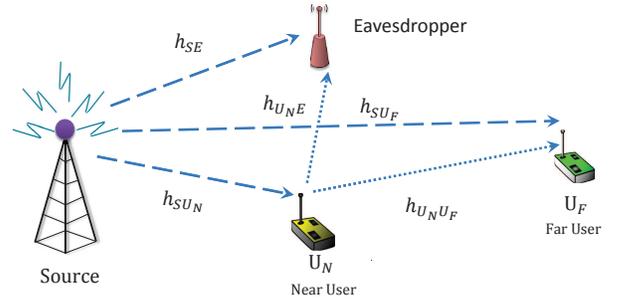}
        \caption{System model.}
    \label{fig:block}
\end{figure}

Let us consider a cooperative relaying system consisting of a source (S), and two destinations ($U_N$ and $U_F$) in the presence of an eavesdropper (E) as shown in Figure \ref{fig:block}. The nodes $U_N$, $U_F$ and E are able to decode information and harvest energy from the received RF signal. It is assumed that $U_N$, $U_F$ and E have the channel state information (CSI) of their corresponding links, whereas, S being the source has the CSI of all the nodes. The channel gains from S $\to U_N$, S $\to U_F$, S $\to$ E and $U_N$ $\to$ E are assumed to be Rayleigh distributed and given as $h_{SU_{N}}$, $h_{SU_{F}}$, $h_{SE}$, $h_{U_{N}E}$, respectively.

The transmission takes place in two time slots. In the first phase, S transmits the superimposed message $\sqrt[]{\alpha 
_{N}P}s_{N}+\sqrt[]{\alpha _{F}P}s_{F}$ to $U_N$ and $U_F$, where $s_{i}$ and $\alpha _{i}$ are the data symbol and power allocation coefficient of $i$-th destination and $P$ denotes the total transmit power. It is assumed that $h_{SU_{N}}>h_{SU_{F}}$, hence the power allocation factor should satisfy $\alpha _{N}<\alpha _{F}$, where $\alpha _{N}+\alpha _{F}=1$. The nodes $U_N$, $U_F$ and E are assumed to use the power-splitting receiver architecture for ID and EH. According to power-splitting architecture, the received power is split into two power streams by a 
power-splitting factor $\rho$ for EH and $(1-\rho )$ for ID, where $0<\rho <1$. The received signal at $U_N$, $U_F$ and E during the first time slot can be written as

\begin{align}
y_{SU_{N}}^{(1)}&=\sqrt[]{(1-\rho _{N,1})}(\sqrt[]{\alpha 
_{N}P}s_{N}+\sqrt[]{\alpha _{F}P}s_{F})h_{SU_{N}}+n_{SU_{N}},\\
y_{SU_{F}}^{(1)}&=\sqrt[]{(1-\rho _{F,1})}(\sqrt[]{\alpha 
_{N}P}s_{N}+\sqrt[]{\alpha _{F}P}s_{F})h_{SU_{F}}+n_{SU_{F}},\\
y_{SE}^{(1)}&=\sqrt[]{(1-\rho _{E,1})}(\sqrt[]{\alpha 
_{N}P}s_{N}+\sqrt[]{\alpha _{F}P}s_{F})h_{SE}+n_{SE},
\end{align}
where $\rho _{N,1}$, $\rho _{F,1}$, $\rho _{E,1}$ denote power-splitting factor at $U_N$, $U_F$ and E during first phase. Also, $n_{SU_{N}}$, $n_{SU_{F}}$, $n_{SE}$ represent the additive white Gaussian noise (AWGN) with zero mean and $N_{0}$ variance. The node $U_N$ first decodes its own symbol $s_{N}$ by treating $s_{F}$ as interference. After obtaining $s_{N}$, $U_N$ cancels its own signal by using successive interference cancellation (SIC) to get $s_{F}$. The received signal to interference and noise ratio (SINR) and signal-to-noise ratio (SNR) for symbols $s_{N}$ and $s_{F}$ can be, respectively, given as

\begin{align}
{\gamma }_{SU_{N}}^{(1)}&=\frac{(1-\rho _{N,1})\vert h_{SU_{N}}\vert 
^{2}\alpha _{F}P}{(1-\rho _{N,1})\vert h_{SU_{N}}\vert ^{2}\alpha 
_{N}P+N_{0}},\\
\gamma _{SU_{N}}^{(2)}&=\frac{(1-\rho _{N,1})\vert h_{SU_{N}}\vert ^{2}\alpha 
_{N}P}{N_{0}}.
\end{align}

The far user, $U_F$, treats the $s_{N}$ as interference. Then the received SINR at $U_F$ can be written as

\begin{align}
{\gamma }_{SU_{F}}^{(1)}=\frac{(1-\rho _{F,1})\vert h_{SU_{F}}\vert 
^{2}\alpha _{F}P}{(1-\rho _{F,1})\vert h_{SU_{F}}\vert ^{2}\alpha 
_{N}P+N_{0}}.
\end{align}

It is assumed that the link between the source and near user is secure and the eavesdropper tries to decode the information signal of the far user. In order to decode $s_{F}$, the eavesdropper treats $s_{N}$ as noise. Hence, the SINR at E can be expressed as

\begin{align}
{\gamma }_{SE}^{(1)}=\frac{(1-\rho _{E,1})\vert h_{SE}\vert ^{2}\alpha 
_{F}P}{(1-\rho _{E,1})\vert h_{SE}\vert ^{2}\alpha _{N}P+N_{0}}.
\end{align}

In the second phase, $U_N$ transmits the decoded symbol $s_{F}$ to $U_F$ with power $P$. Assuming that $U_N$ can perfectly decode $s_{F}$and use all harvested energy $\frac{E_{N}}{T}=\rho _{N,1}\eta P\vert h_{SU_{N}}\vert ^{2}$ during first phase to transmit $s_{F}$ to $U_F$, the received SNR at $U_F$ can be given as

\begin{align}
\gamma _{U_{N}U_{F}}^{(2)}=\frac{(1-\rho _{F,2})\rho _{N,1}\eta \vert 
h_{SU_{N}}\vert ^{2}\vert h_{U_{N}U_{F}}\vert ^{2}P}{N_{0}}.
\end{align}

Similarly, the received SNR at E during the second phase can be given as

\begin{align}
\gamma _{U_{N}E}^{(2)}=\frac{(1-\rho _{E,2})\rho _{N,1}\eta \vert 
h_{SU_{N}}\vert ^{2}\vert h_{U_{N}E}\vert ^{2}P}{N_{0}}.
\end{align}

\section{Intercept Probability}

In this section, we derive the analytical expression of intercept probability for the considered case. An intercept event occurs when the 
achievable secrecy rate $C_{\sec }$ falls below 0 \cite{8103768}. The achievable secrecy rate is the difference between the rates of the main and wiretap links i.e. $C_{\sec }=\lbrack C_{s}-C_{e}\rbrack ^{+}$.

Since we consider DF protocol at $U_N$, therefore, the SNR will be determined by the bottleneck link between S and $U_F$ and $U_N$ to 
$U_F$ is given as $\gamma _{S}^{DF}=\min({\gamma }_{SU_{F}}^{(1)},\gamma _{U_{N}U_{F}}^{(2)})$ while the achievable rate for the link S $\to U_N \to U_F$ is given as $C_{s}^{DF}=\frac{1}{2}\log _{2}(1+\gamma _{S}^{DF})$. The eavesdropper is assumed to select the best messages received during first and second phase given as $\gamma _{E}^{DF}=\max ({\gamma}_{SE}^{(1)},\gamma _{U_{N}E}^{(2)})$. Therefore, the achievable rate 
for wiretap links can be given as $C_{e}=\frac{1}{2}\log _{2}(1+\gamma _{E}^{DF})$. Now the intercept probability can be given as

\begin{align}
P_{int}^{DF}=\Pr\left\{ \frac{1}{2}\log _{2}\left(\frac{1+\min ({\gamma 
}_{SU_{F}}^{(1)},\gamma _{U_{N}U_{F}}^{(2)})}{1+\max ({\gamma 
}_{SE}^{(1)},\gamma _{U_{N}E}^{(2)})}\right)<0\right\}.
\label{eq_1}
\end{align}

Assuming $X=\min ({\gamma }_{SU_{F}}^{(1)},\gamma _{U_{N}U_{F}}^{(2)})$ and $Y=\max ({\gamma }_{SE}^{(1)},\gamma 
_{U_{N}E}^{(2)})$, we can re-write \ref{eq_1} as 
%%%%%%%%%%%%%%%%%%%%%%%%%%%%%5
\begin{align}
P_{int}^{DF}&=\Pr\lbrack X<Y\rbrack \nonumber \\
&=\int_{0}^{\infty }{F_{X}(y)}f_{Y}(y)dy.
\label{eq_2}
\end{align} 
%%%%%%%%%%%%%%%%%%%%%%%%%%%%%%

The Cumulative Distribution Function (CDF) can be found as

\begin{align}
F_{X}(x)=1-(1-F_{{\gamma }_{SU_{F}}^{(1)}}(x))(1-F_{\gamma 
_{U_{N}U_{F}}^{(2)}}(x)).
\label{eq_3}
\end{align}

In the above equation, we obtain

\begin{align}
F_{{\gamma }_{SU_{F}}^{(1)}}(x)&=\Pr\left(\frac{(1-\rho _{F,1})\vert 
h_{SU_{F}}\vert ^{2}\alpha _{F}P}{(1-\rho _{F,1})\vert h_{SU_{F}}\vert 
^{2}\alpha _{N}P+N_{0}}<x\right)\nonumber \\
&=1-\exp\left\{ -\frac{x}{\Omega _{SU_{F}}(1-\rho _{F,1})(\alpha 
_{F}-\alpha _{N}x)}\right\} \nonumber \\
&=1-\frac{\Gamma \left(m,\lambda \frac{x}{(1-\rho _{F,1})(\alpha _{F}-\alpha 
_{N}x)}\right)}{\Gamma (m)},
\end{align}
where $\lambda _{SU_{F}}=\frac{m}{\Omega _{SU_{F}}}$ and $\Omega _{SU_{F}}=\frac{PE\lbrace \vert h_{SU_{F}}\vert ^{2}\rbrace }{N_{0}}$. The CDF of $\gamma _{U_{N}U_{F}}^{(2)}$ depends on the event that near user has successfully decoded the symbol of the far user. In this case, the CDF of $\gamma _{U_{N}U_{F}}^{(2)}$ can be expressed as

\begin{align}
F_{\gamma _{U_{N}U_{F}}^{(2)}}(x)=\Pr\biggl(\frac{(1-\rho _{N,1})\vert 
h_{SU_{N}}\vert ^{2}\alpha _{F}P}{(1-\rho _{N,1})\vert h_{SU_{N}}\vert 
^{2}\alpha _{N}P+N_{0}}>x \nonumber \\
,\frac{(1-\rho _{F,2})\rho _{N,1}\eta \vert 
h_{SU_{N}}\vert ^{2}\vert h_{U_{N}U_{F}}\vert ^{2}P}{N_{0}}<x\biggr).
\end{align}

Assuming $X_{1}=\vert h_{SU_{N}}\vert ^{2}$and $X_{2}=\vert 
h_{U_{N}U_{F}}\vert ^{2}$ we get

\begin{align}
F_{\gamma _{U_{N}U_{F}}^{(2)}}(x)&=\underbrace{\int_{\Theta}^{\infty 
}{f_{X_{1}}(x_{1})dx_{1}}}_{\Psi 
_{1}} \nonumber \\
&-\underbrace{\int_{\Theta}^{\infty }{\bar{F}_{X_{2}}\left(\frac{N_{0}x}{(1-\rho _{F,2})\rho 
_{N,1}\eta Px_{1}}\right)f_{X_{1}}(x_{1})dx_{1}}}_{\Psi _{2}},
\end{align}

where $\bar{F}(.)$ is the complementary CDF, $\Theta=\frac{x}{(1-\rho _{N,1})(\alpha _{F}-\alpha 
_{N}x)}$, and $\Psi _{1}=\frac{\Gamma (m,\lambda \Theta)}{\Gamma (m)}$.

\begin{align}
\Psi _{2}&=\lambda _{SU_{N}}^{m}\int_{\Theta}^{\infty }{\sum_{s=0}^{m-1}{\frac{1}{s!\Gamma 
(m)}\left(\frac{\lambda _{U_{N}U_{F}}x}{(1-\rho _{F,2})\rho _{N,1}\eta 
}\right)^{s}}}\frac{1}{(x_{1})^{s+1-m}} \nonumber \\
&\times \exp\biggl(-\frac{(\lambda _{SU_{N}}(1-\rho _{F,2})\rho _{N,1}\eta 
(x_{1})^{2})}{(1-\rho 
_{N,1})\alpha _{N}(1-\rho _{F,2})\rho _{N,1}\eta x_{1}} \nonumber \\
&-\frac{(\lambda _{U_{N}U_{F}}x(1-\rho _{N,1})\alpha _{N})}{(1-\rho 
_{N,1})\alpha _{N}(1-\rho _{F,2})\rho _{N,1}\eta x_{1}}
\biggr)dx_{1}
\end{align}

\begin{align}
&\Psi _{2}=(-1)^{s-m+1}\left\{\frac{\lambda _{SU_{N}}(1-\rho _{F,2})\rho _{N,1}\eta 
}{(1-\rho _{N,1})\alpha _{N}(1-\rho _{F,2})\rho _{N,1}\eta 
}\right\}^{s-m} \nonumber \\
&\times Ei\left(-\frac{\lambda _{SU_{N}}(1-\rho _{F,2}) 
x}{(1-\rho _{N,1})\alpha _{N}(1-\rho _{F,2})(1-\rho _{N,1})(\alpha 
_{F}-\alpha _{N}x)}\right) \nonumber \\
&\times \frac{1}{(s-m)!}+\left(\frac{\lambda _{SU_{N}}(1-\rho _{F,2})\rho _{N,1}\eta }{(1-\rho 
_{N,1})\alpha _{N}(1-\rho _{F,2})\rho _{N,1}\eta }\right)^{k} \nonumber \\
&\times \left(\frac{x}{(1-\rho 
_{N,1})(\alpha _{F}-\alpha _{N}x)}\right)^{k} \frac{\exp\left(-\frac{\lambda _{SU_{N}}x}{(1-\rho _{N,1})\alpha _{N}}\right)}{\left(\frac{x}{(1-\rho _{N,1})(\alpha _{F}-\alpha 
_{N}x)}\right)^{s-m}}\nonumber \\
&\times \sum_{k=0}^{s-m-1}{\frac{(-1)^{k}}{(s-m)(s-m-1)\ldots 
(s-m-k)}}.
\end{align}

Now we find $F_{Y}(y)=\max ({\gamma }_{SE}^{(1)},\gamma 
_{U_{N}E}^{(2)})$ which is given as

\begin{align}
F_{Y}(y)=F_{{\gamma }_{SE}^{(1)}}(y)F_{\gamma _{U_{N}E}^{(2)}}(y)
\end{align}

where $F_{{\gamma }_{SE}^{(1)}}(y)=1-\exp\lbrack -\frac{y}{\Omega _{SE}(1-\rho _{E,1})(\alpha _{F}-\alpha _{N}y)}\rbrack$ and $F_{\gamma _{U_{N}E}^{(2)}}(y)=1-\frac{y}{\Omega _{SU_{N}}\Omega_{U_{N}E}(1-\rho _{E,2})\rho _{N,1}\eta }$. Now by differentiating $F_{Y}(y)$ we obtain

\begin{align}
f_{Y}(y)=\Phi _{1}\left[-1+\exp\left\{ -\frac{y}{\Omega _{SE}(1-\rho _{E,1})(\alpha _{F}-\alpha _{N}y)}\right\} \right] \nonumber \\
+\frac{\alpha _{F}(1-\Phi _{1}y)\exp\left\{ -\frac{y}{\Omega _{SE}(1-\rho _{E,1})(\alpha _{F}-\alpha _{N}y)}\right\} }{\Omega _{SE}(1-\rho _{E,1})(\alpha _{F}-\alpha _{N}y)^{2}},
\label{eq_4}
\end{align}

where $\Phi _{1}=\frac{1}{\Omega _{SU_{N}}\Omega _{U_{N}E}(1-\rho _{E,2})\rho _{N,1}\eta }$ and $\Phi _{2}=\frac{1}{\Omega _{SE}(1-\rho_{E,1})(\alpha _{F}-\alpha _{N}y)}$. By replacing (\ref{eq_3}) and (\ref{eq_4}) in (\ref{eq_2}), we note that a closed-form solution for the integral is mathematically intractable. Still, the intercept probability can be easily obtained by solving the single integral using any computational software.

\begin{figure}
    \centering
        \includegraphics[trim={0 6cm 0 0},clip,width=0.5\textwidth]{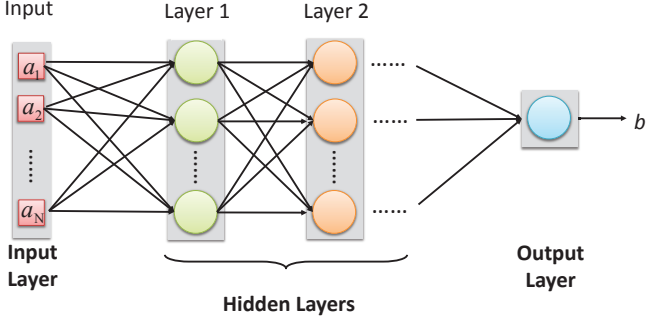}
        \caption{A typical deep neural network with input, output and multiple hidden layers.}
    \label{figblock}
\end{figure}

\section{Deep Learning Based Optimization}

In this section, we are going to present a deep learning based resource allocation scheme for optimizing the achievable secrecy rate of the far user. We employ neural networks to learn the relationship between inputs and outputs and predict the optimal power allocation factor that maximizes the achievable secrecy rate. To do so, we carefully train our multi-layer artificial neural network, whereby, each layer consists of multiple neurons as illustrated in Figure \ref{figblock}. We show in the numerical results section that the computational efficiency of artificial neural networks is one of the highlights of deep learning models. After the model has been trained on a set of inputs, the testing (i.e., the real-time running phase) involves only nonlinear transformations and vector multiplications without compromising the performance. 

\subsection{Problem Formulation}

We now try to optimize the secrecy performance of the far user due to it being most vulnerable to eavesdropping attack. Considering the worst-case scenario\footnote{We consider that the CSI of the eavesdropper is not available at the source due to which the achievable rate of eavesdropper cannot be calculated.}, the legitimate receivers have no option but to maximize their own achievable rate. By this approach, maximizing the achievable rate would result in maximizing the secrecy rate as well since the secrecy rate is the difference between the rate of legitimate link and the rate of wiretap link. Under this condition, the optimization problem of the achievable rate becomes equivalent to 

\begin{align}
\max_{\alpha_F>0.5} C_{sec} \equiv \max_{\alpha_F>0.5}  \frac{1}{2}\log _{2}\left(\min ({\gamma 
}_{SU_{F}}^{(1)},\gamma _{U_{N}U_{F}}^{(2)})\right).
\label{eqrq}
\end{align} 

However, the second term (i.e., $\gamma _{U_{N}U_{F}}^{(2)}$) in (\ref{eqrq}) does not contain power allocation factor $\alpha_F$. Thus, the optimization problem can be re-formulated as $\max_{\alpha_F>0.5} \log_2(1+{\gamma }_{SU_{F}}^{(1)})$. 

\subsection{Deep Learning Network Setup}

Our neural network consists of multiple hidden layers and a single input and output layer. The main reason for using multiple hidden layers is to avoid under-fitting of test data while maintaining a sufficient level of complexity. Moreover, by utilizing multiple hidden layers, the complex interplay of inputs and outputs can be understood by the network during the learning phase. In our case, the inputs are the channel realizations and the outputs are the power allocation factors of the far user. For each channel realization, we take samples from the Rayleigh distribution while fixing all the other parameters. These values are generated for training and validation datasets that are fed into the network during the training phase.

At each hidden layer, we use rectified linear unit (ReLU) activation function. Mathematically, the ReLU activation function is represented as

\begin{align}
z=\max(y,0), 
\end{align}
where $z$ denotes the output of the activation function and $y$ is the input of the function. We have used mean square error as the cost function and apply the mini-batch algorithm on the training data samples for calculating the gradients. 

\section{Numerical Results}

This section provides numerical results and relevant discussion. It is worth mentioning that the analytical and simulation results have been generated using MATLAB, whereas, the deep learning optimization is performed in Python 3.6.7. Unless stated otherwise, the parameters used for generation of plots are as follows: $\rho _{N,1}=\rho _{N,2}=\rho _{E,1}=\rho _{E,2}=\rho _{F,1}=\rho _{F,2}=\rho=0.3$, $\Omega=5$dB, decay rate = 0.9, training samples = 30000, test samples= 6000, and epochs=100. 

\begin{figure}
    \centering
        \includegraphics[trim={0 0cm 0 0},clip,width=0.45\textwidth]{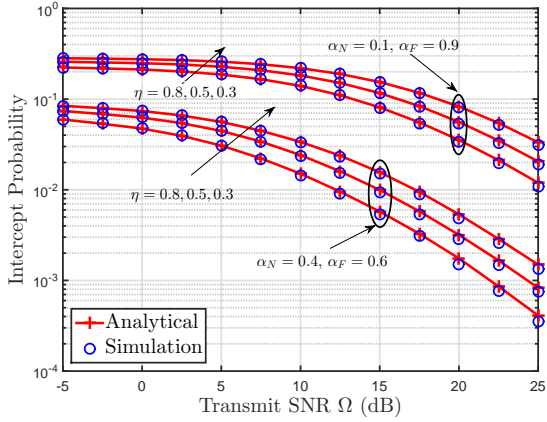}
        \caption{Intercept probability versus transmit SNR.}
    \label{fig11}
\end{figure}

Figure \ref{fig11} illustrates the intercept probability against different values of transmit SNR. It can be seen that the intercept probability decreases with an increase in the values of $\Omega$. As anticipated, the values of $\eta$ also have a prominent impact on the intercept probability. Strictly speaking, the intercept probability generally increases with a reduction in the values of $\eta$. However, the separation between the curves of $\eta$ grows as the value of transmitting SNR increases. This shows that a low energy harvesting efficiency is more harmful at higher values of SNR, giving rise to a higher intercept probability. In addition to this, we note that at higher values of $\alpha_F$, the intercept probability significantly increases. This is partly because of the low power allocation to the near cooperative user. At lower values of the power allocation factor, decoding the message of the far user becomes difficult for the near cooperating user. It can also be seen that the simulation results closely follow the analytical result which validates the derived expression.

\begin{figure}
    \centering
        \includegraphics[trim={0 0cm 0 0},clip,width=0.45\textwidth]{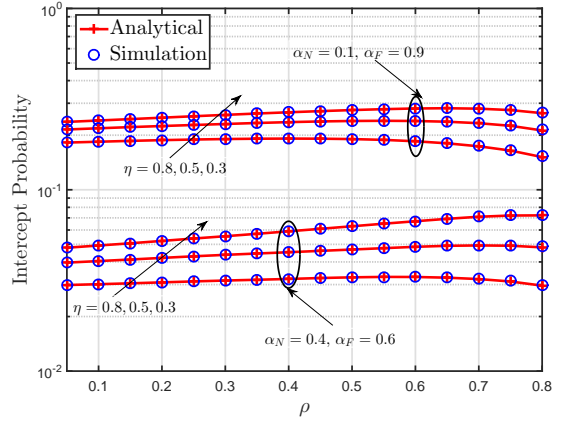}
        \caption{Intercept probability against power-splitting factor.}
    \label{fig22}
\end{figure}

To further highlight the impact of a power-splitting factor, Figure \ref{fig22} shows the intercept probability as a function of $\rho$. It can be observed that the intercept probability generally increases with an increase in the value of the power-splitting factor. This trend can be attributed to the low amount of energy reserved for information decoding which makes it difficult to maintain cooperation among near and far users. Evidently, the increasing values of $\eta$ causes intercept probability to decreases. However, the power allocation factors for near and far users have shown different trends as the value of $\rho$ changes. Precisely, we note that at lower values of power-splitting factors, the separation between the curves of $\alpha_F=0.9$ and $\alpha_F=0.6$ is quite large. But, as the values of the power-splitting factor increase, the difference between the curves becomes smaller. This shows that the impact of power allocation factors of NOMA reduces at higher values of a power-splitting factor.   

\begin{figure}
    \centering
    \begin{tabular}{cc}
        \includegraphics[trim={0 0cm 0 0},clip,width=0.35\textwidth]{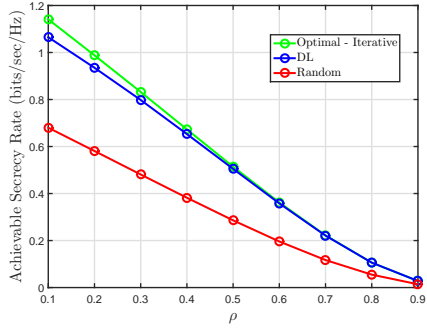} \\
        (a) \\
        \includegraphics[trim={0 0cm 0 0},clip,width=0.35\textwidth]{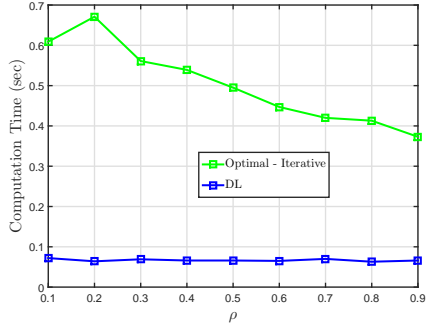} \\
        (b)
        \end{tabular}
        \caption{Performance comparison (a) achievable secrecy rate versus $\rho$ (b) Computation time versus $\rho$. The neural network contain 02 hidden layers: Layer 1 = 200 neurons, and Layer 2 = 100 neurons. \vspace{-2cm}}
    \label{fig_two}
\end{figure}

In Figure \ref{fig_two}, we have demonstrated the optimization results for the deep learning approach. Here, ``Optimal-Iterative'' refers to the optimal results achieved through iterative search scheme, ``DL'' denotes the results for deep learning approach, and ``Random'' represents the results for random power allocation factor generated using uniform distribution. Figure \ref{fig_two}(a) shows the results for achievable secrecy rate against the increasing values of the power-splitting factor. As shown in this plot, the larger values of the power-splitting factor reduce the achievable secrecy capacity as more power is reserved for energy harvesting. It can be seen that the deep learning approach strictly follows the optimal results, while always achieving the accuracy of more than 90\%. By contrast, the random power allocation achieves a very low secrecy capacity. To further highlight the robustness of the deep learning approach, Figure \ref{fig_two}(b) plots the results for computation time against different values of $\rho$. For this case, the deep learning approach performs significantly better than that of iterative search. This shows that once trained, the deep learning models can provide a lightweight solution to achieve optimal results.  

\begin{figure}
    \centering
    \begin{tabular}{c}
        \includegraphics[trim={0 0cm 0 0},clip,width=0.35\textwidth]{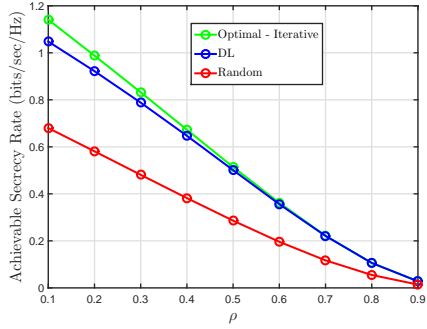} \\
        (a) \\
        \includegraphics[trim={0 0cm 0 0},clip,width=0.35\textwidth]{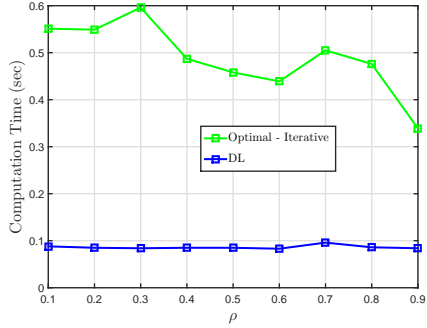} \\
        (b)
        \end{tabular}
        \caption{Performance comparison (a) achievable secrecy rate versus $\rho$ (b) Computation time versus $\rho$. The neural network contain 05 hidden layers: Layer 1 = 200 neurons, Layer 2 = 100 neurons, Layer 3 = 50 neurons, Layer 4 = 30 neurons, and Layer 5 = 10 neurons. }
    \label{fig_five}
\end{figure}

Figure \ref{fig_five} emphasizes the suitable number of layers for the neural network but plotting achievable secrecy rate and computation time for 05 hidden layers. It can be seen in Figure \ref{fig_five}(a) that the separation between deep learning and iterative approach slightly increases. We attribute this increase to over-fitting of data during the training phase. This results in causing the neural network to memorize the training set. However, when new testing data is presented, the network fails to generalize the results. This increase in a number of hidden layers also affects the computation time, as shown in Figure \ref{fig_five}(b). Specifically, we observe that the computation time of deep learning approach considerably increases as compared to Figure \ref{fig_two}(b). This increase in computation time is due to the increase in the number of hidden layers and in the total number of neurons in the neural network. 
\vspace{-0.2cm}
\section{Conclusion}

This study provides secrecy analysis and deep learning optimization of SWIPT-based cooperative NOMA systems. We derive the analytical expression of the intercept probability when near user act as a cooperative node in the presence of an eavesdropper. We have shown that the impact of power allocation factors of NOMA reduces at higher values of the power-splitting factor. Moreover, we have shown that deep learning approach is more robust and computationally efficient as compared to conventional iterative search approach. In the future, we aim to use deep learning for optimizing the secrecy performance of cooperative NOMA systems under colluding eavesdroppers. 
\vspace{-0.3cm}
\section*{Acknowledgment}
This work is partially supported by the National Key R \& D Plan (2017YFC0803403), the National Natural Science Foundation of China (61371188) and the Fundamental Research Funds of Shandong University (2018GN051).

\bibliographystyle{IEEEtran}% This is IEEEtran.bst file

\bibliography{References}

\end{document}